\documentclass[a4paper]{PoS}
\usepackage{subfig}

\title{Measurement of shower development and its \textit{Moli\`ere radius} with a four-plane LumiCal test set-up}

\ShortTitle{Measurement of shower development and its \textit{Moli\`ere radius} with a four-plane LumiCal test set-up}

\author{\speaker{Itamar Levy}\\
On behalf of the FCAL collaboration.\\
       Tel Aviv University (IL)\\
       E-mail: \email{itamar@alzt.tau.ac.il}}

\abstract{A prototype of a luminometer, designed for a future $e^{+}e^{-}$ collider detector, and consisting at present of a four-plane module, was tested in the CERN PS accelerator T9 beam. The objective of this beam test was to demonstrate a multi-plane operation, to study the development of the electromagnetic shower and to compare it with MC simulations. In addition, the effective \textit{Moli\`ere radius} of this configuration is extracted.}

\FullConference{The European Physical Society Conference on High Energy Physics\\
                 5-12 July\\
                 Venice, Italy}

\begin{document}

\section{Introduction}                
\begin{figure}[h!]
\begin{center}
\subfloat[\label{fig:Forward_structure}]{\includegraphics[width=.49\textwidth]{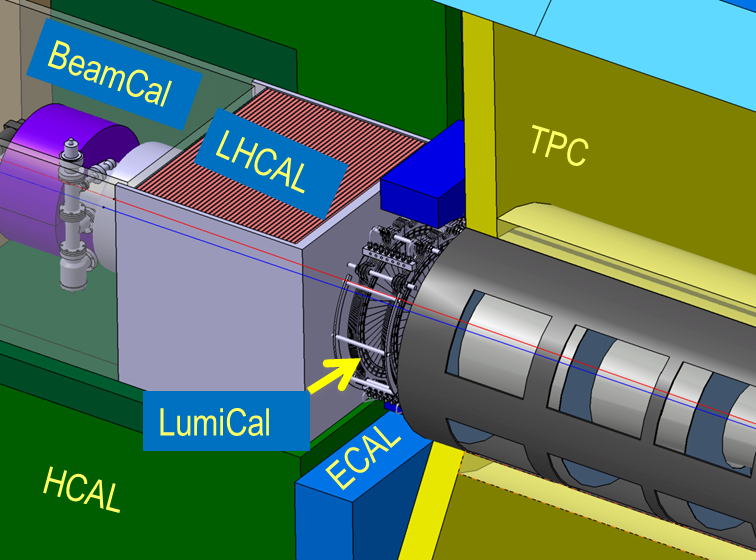}}
\hfill
\subfloat[\label{fig:sensor}]{\includegraphics[width=.30\textwidth]{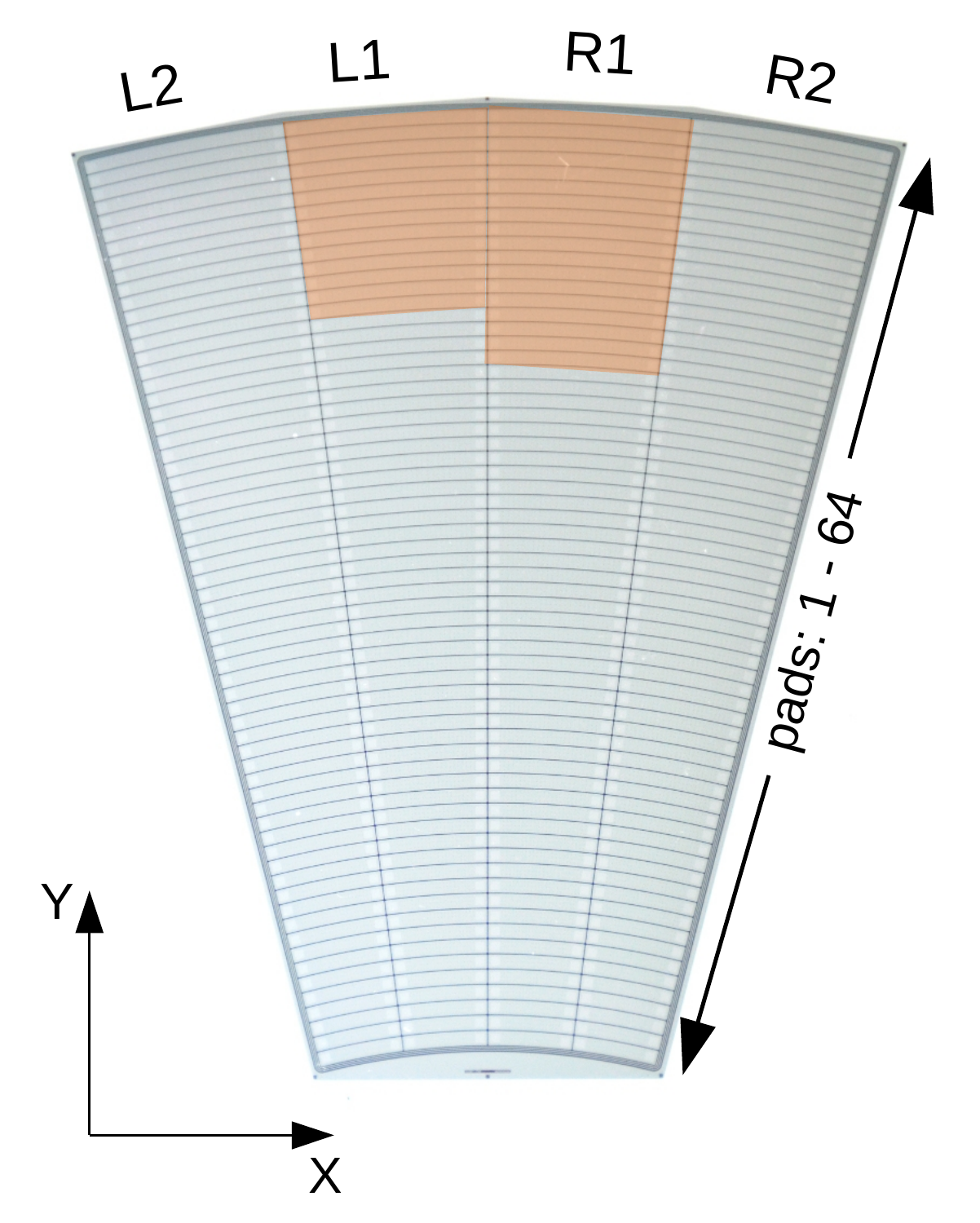}}
\caption{(a) The very forward region of the ILD detector. (b) A LumiCal prototype silicon sensor. The shaded area denotes the pads that were connected in this test. }
\label{fig:Forward_structure_sensor}
\end{center}
\end{figure}

The development and optimization of the instrumentation in the very forward region of a detector for a future $e^+ e^-$ linear collider is done by the FCAL collaboration.
The forward region main goals are to  measure the bunch by bunch and the precise integrated luminosity, to give a fast feedback for beam monitoring and tuning, and to extend the detector coverage to small polar angles. 
In order to meet the forward region goals, two compact electromagnetic calorimeters are foreseen in the forward region~\cite{cite:fcal_jinst}.
The Luminosity Calorimeter, LumiCal, is designed to measure the precise integrated luminosity  using the rate of low angle Bhabha events.
The Beam Calorimeter, BeamCal, will perform a bunch-by-bunch estimate of the luminosity and assist in beam tuning.
A sketch of the forward region layout of the ILD detector is shown in Figure~\ref{fig:Forward_structure}.
Due to the requirements in the forward region the calorimeters must have fast readout, radiation hard sensors and a strict mechanical precision.
Both calorimeters are foreseen as sampling calorimeters, with tungsten as an absorber.
The LumiCal design uses silicon as the sensitive layer, while the BeamCal base design uses GaAs.

Since the luminosity precision depends on the granularity of the radial direction, LumiCal sensitive layers are segmented as long and narrow pads.
In order to meet the required precision, the LumiCal design has 64 radial pads in each sector with a 1.8~mm pitch.
The  azimuthal direction is divided into 48 sector, each $7.5^{\circ}$ in angle.
High resistivity silicon sensors prototype were produced by Hamamatsu from $6^{\prime\prime} $ wafers.
Each sensor tile is $320~\mu m$ thick, and contains 4 sectors ($30^{\circ}$ in total).
Pads are made of $p^{+}$ implants in $n^{-}$ type bulk, and DC coupled with read out electronics.
One sensor tile can be seen at Figure~\ref{fig:sensor}.

The performance of fully instrumented LumiCal and BeamCal detector planes was studied in
previous beam test campaigns \cite{cite:TB11}. The next step in the detector prototype development was to preform a beam test study of a multi-plane
structure. A brief summary of the first LumiCal prototype test results is present here. 

\section{The 2014 test beam}
\begin{figure}[!ht]
	\centering
	\includegraphics[width=0.95\columnwidth]{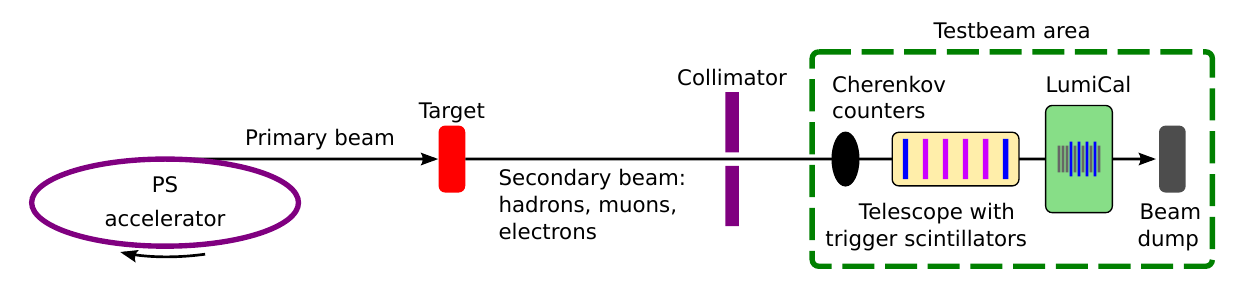}
	\caption{Schematic view of the beam and experimental set-up (not to scale).}
	\label{fig:tb_beam}
\end{figure}

The first multi-layers test was performed in October 2014 at the T9 East area of the proton synchrotron (PS) at CERN.
The PS accelerator primary proton beam hits a target, producing the secondary beam to the T9 area which consists of a mixture of electrons, muons and hadrons.
A narrow band of 5 GeV particle momenta was selected.
A four planes pixel telescope was used to measure the trajectories of beam particles. 
The telescope utilizes MIMOSA-26 chips~\cite{mimosa26} developed by the Aarhus University in collaboration with the Strasbourg University.
Hit resolution of $9~\mu m$ was determined~\cite{cite:Oron} in each plane of the telescope.
Three scintillation counters were used to provide a trigger for particles traversing the active part of the telescope sensors.
The trigger signal was combined with the Cherenkov counters response to create a trigger for leptons. 
The simplified overall view of the beam and the experimental set-up is presented in Figure~\ref{fig:tb_beam}.

To allow the multiple-plane operation, a mechanical structure to meet the demanding geometrical
requirements was developed~\cite{nuiry}. 
The sensors were mounted onto 2.5 mm thick PCB serving as the mechanical support and high voltage supply.
The thickness of the sensor board force to dedicate an absorber slot to each sensor forming a gap of $5.5~mm$. 
Since only four readout boards were available, three detector configurations were used, with the active sensor layers always separated by two absorber layers.
By adding additional absorber layers upstream of the detector, the sensor layers were effectively moved downstream in the shower. 
As shown in Figure~\ref{fig:sensor} pads 51-64 of sector L1 and 47-64 of sector R1 (32 channels in total) were connected in each sensor.
Each sensor was read out by one electronic board~\cite{fcal_readout_board}, containing 4 pairs of dedicated, 8 channels, front end and 10-bit pipeline ADC ASICs.

\section{Test beam results}
Since the beam particles arrive stochastically, the LumiCal readout module utilizes an asynchronous sampling mode.
First, an initial treatment of each waveform that includes the baseline and the common-mode subtraction, was performed, 
followed by reconstruction of signal amplitudes using a deconvolution procedure~\cite{Jakub}.

In order to use the same energy scale in both data and simulation, the response for beam muons  for each layer was used. 
The value of the most probable value of muon depositions was defined as one unit of minimum ionizing particle (MIP).
This value was used to scale all energy measurements, both in the data analysis and in the GEANT4 Monte Carlo (MC) simulation.
In the data, this calibration procedure, imposed an estimated 5\% uncertainty.

\subsection{longitudinal shower development}
\begin{figure}[h!]
\begin{center}
\subfloat[\label{fig:shower_full_1}]{\includegraphics[width=.45\textwidth]{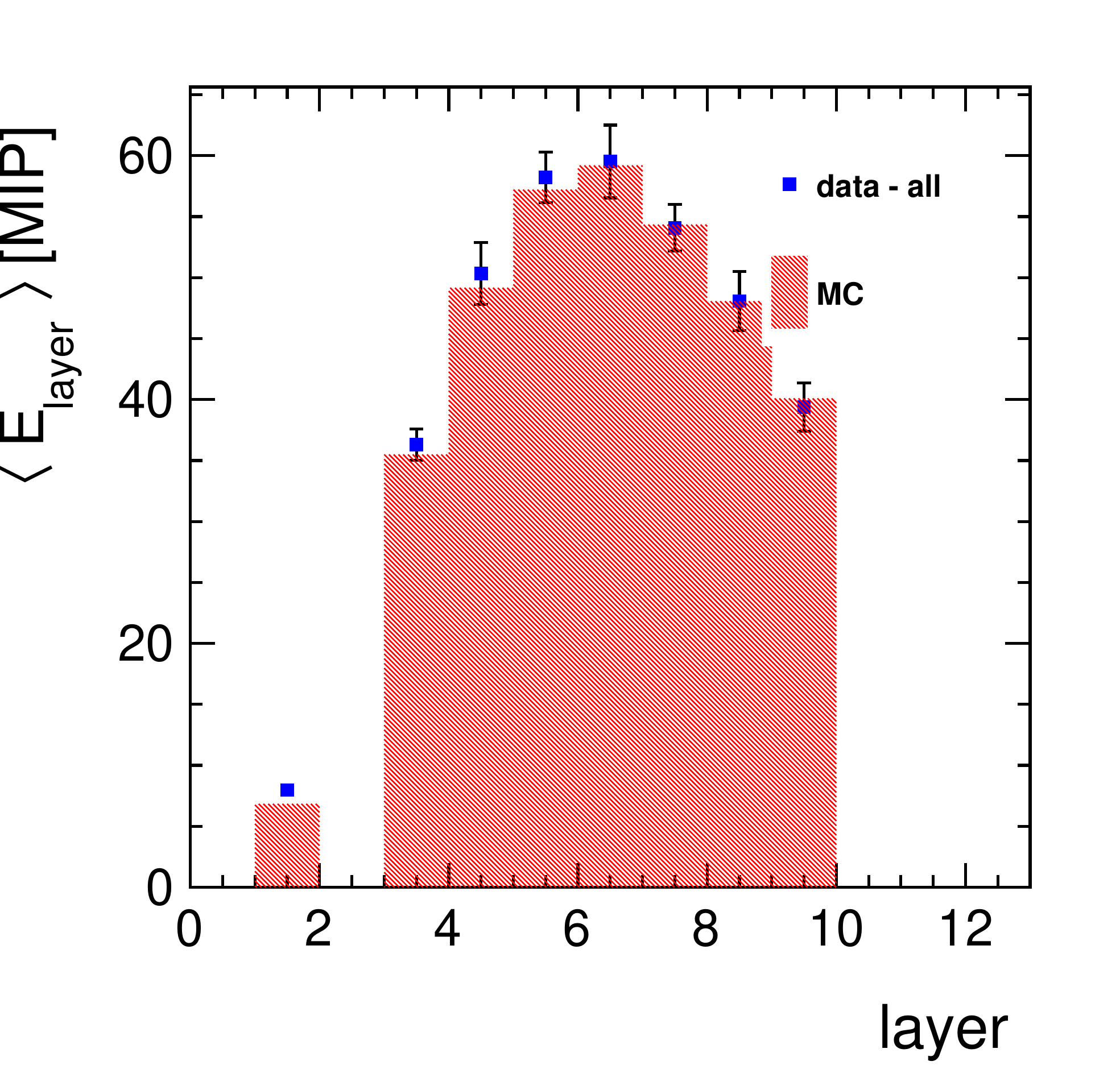}}
\hfill
\subfloat[\label{fig:shower_full_2}]{\includegraphics[width=.45\textwidth]{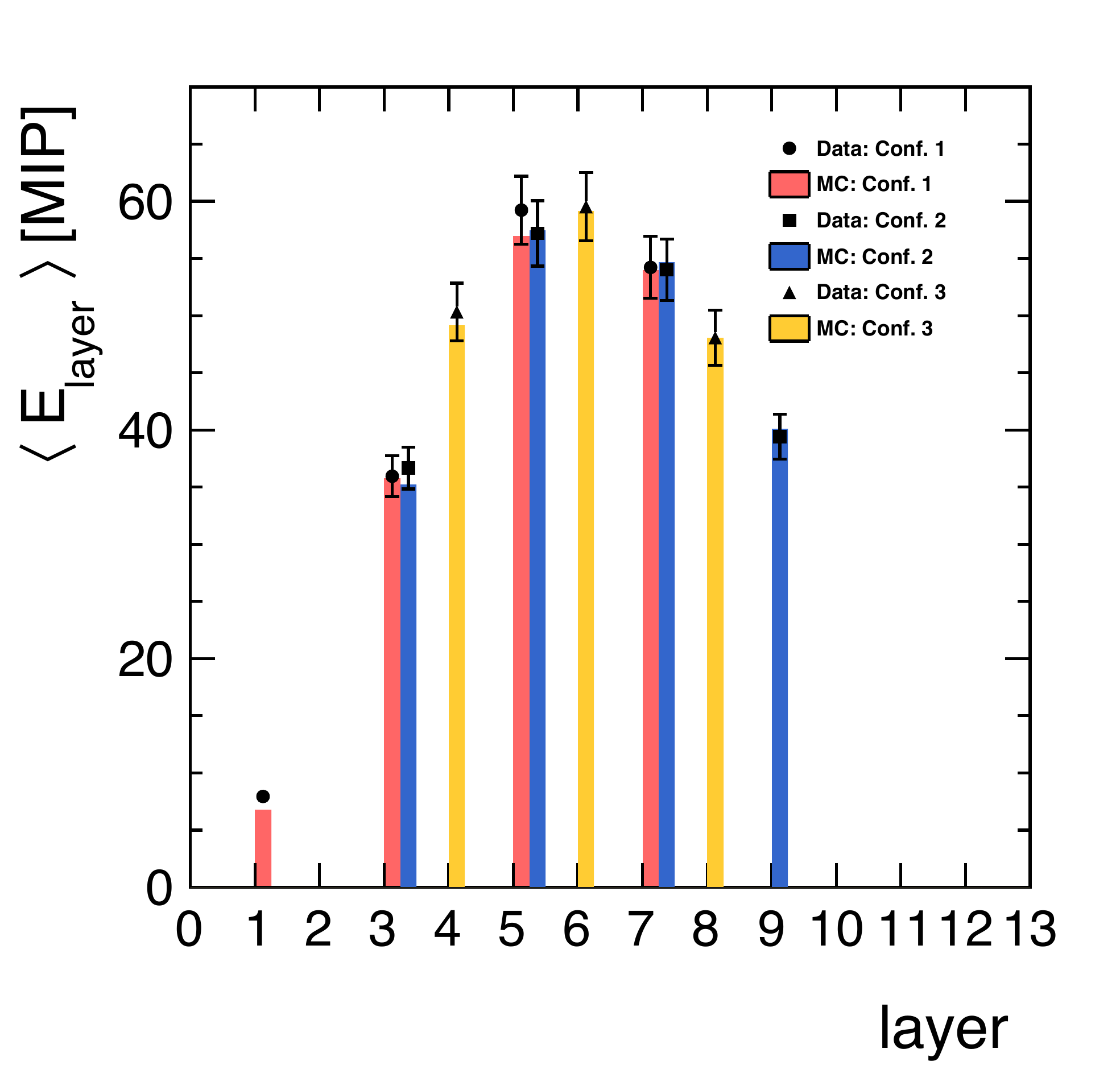}}
\caption{Average energy deposited in the detector planes as a function of the number of absorber layers for three configurations combined (a) and
separately (b). The dots are data and the shaded area corresponds to the MC simulation.}
\label{fig:shower_full}
\end{center}
\end{figure}

The longitudinal development of electron showers is shown in Figure~\ref{fig:shower_full_1} in terms of average
shower energy deposits per plane as a function of the number of absorber layers. In Figure~\ref{fig:shower_full_2} the
development according to the three different configurations are presented separately, each color represents
one configuration. Here the common layers in different configurations can be compared and are in
good agreement with each other. The results were compared with the prediction of the simulation,
and agreement between the simulation and the data is found within the uncertainties.

The difference in layer 1, where the simulated deposition is slightly smaller, is understood as due to preshowering caused by upstream elements.
The uncertainty on the data is dominated by a 5\% calibration uncertainty. 

\subsection{Position resolution}

\begin{figure}[h!]
\begin{center}
\subfloat[\label{fig:position_1}]{\includegraphics[width=.45\textwidth]{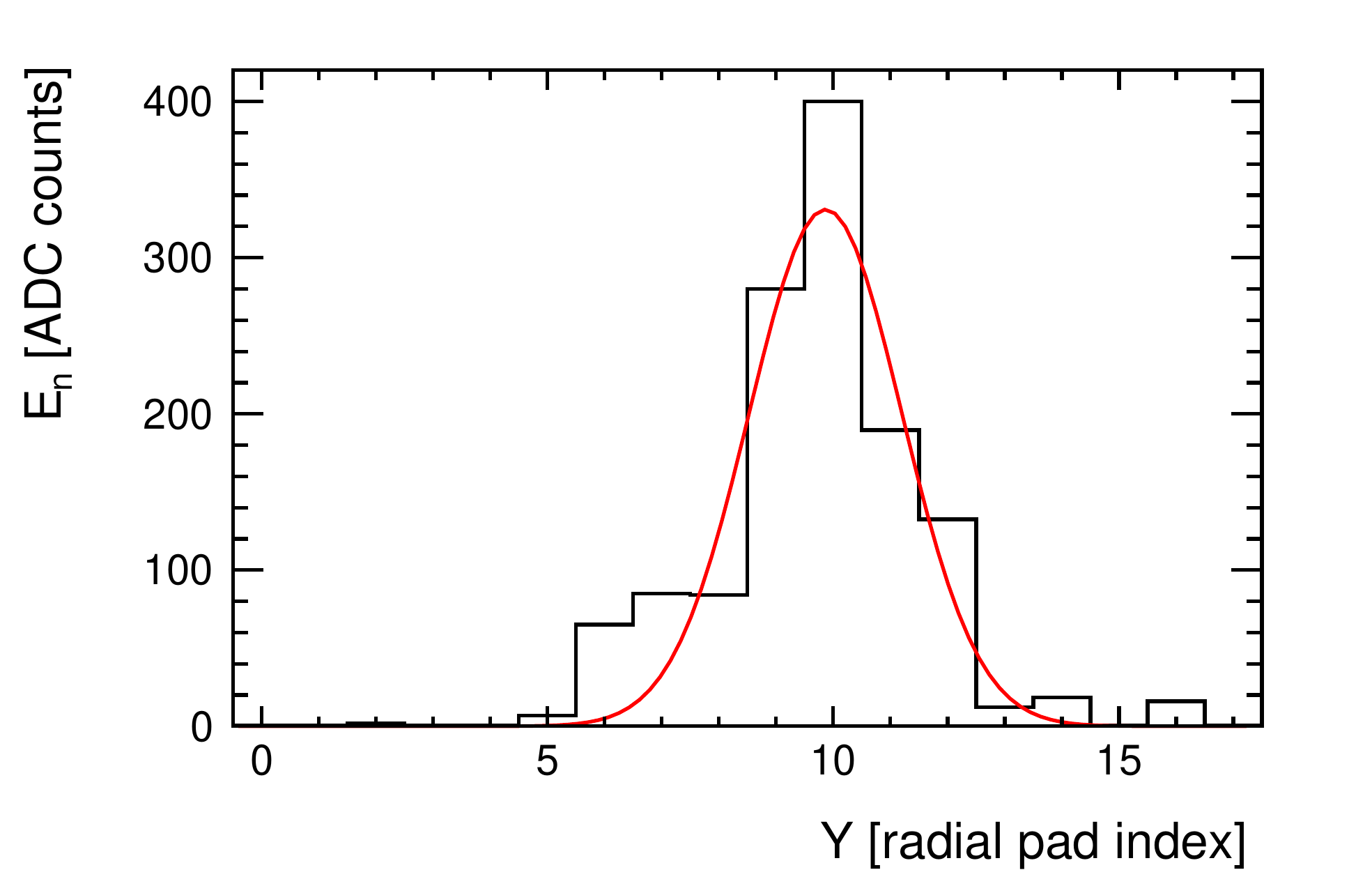}}
\hfill
\subfloat[\label{fig:position_2}]{\includegraphics[width=.45\textwidth]{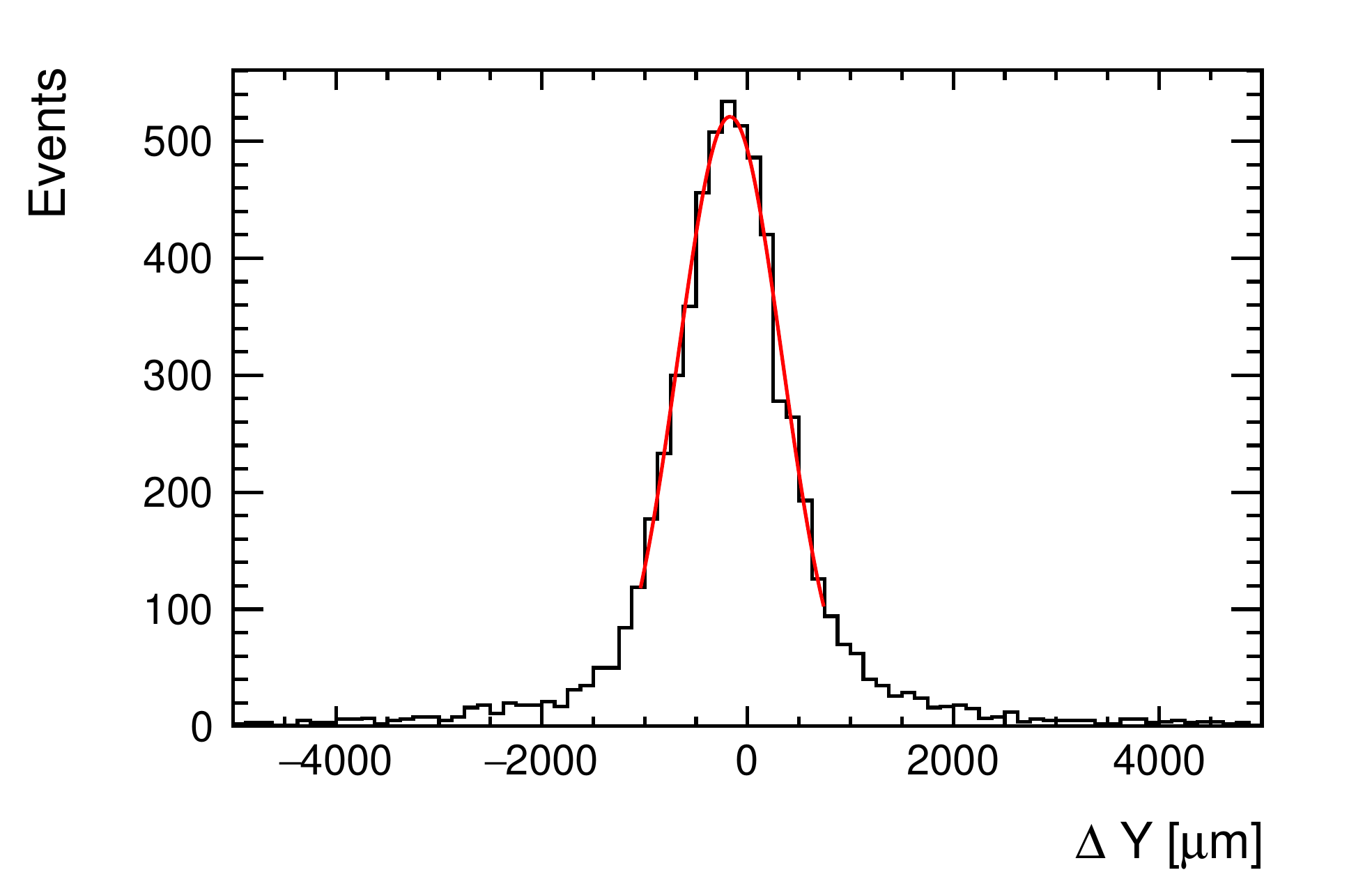}}
\caption{(a) The deposited energy sum for a single event, $E_n$ , as a function of Y expressed in terms
of the pad number. The curve is a Gaussian fit to the data. (b) The distribution of the residuals between the reconstructed positions, $\Delta Y$. The curve represents the Gaussian fit to the distribution }
\label{fig:position}
\end{center}
\end{figure}

The energy deposited in the sensor pad for layer $l$, sector $k$ and radial pad index $n$ is denoted as  $E_{nkl}$. 
The one-dimensional deposited energy distribution for one event along the radial axis Y can be obtained from the energy of a tower. 
The energy of a tower at the pad index $n$ is the sum over the layer index $l$ and the sector index $k$, and is denoted by $E_n$. 
An example of the $E_n$ distribution for a single event is shown in Figure~\ref{fig:position_1}.
The position of the hit on the surface of the first layer projected to the radial coordinate Y can be estimated as the mean of the Gaussian fitted to the $E_n$ distribution.

For each event, the reconstructed position of the shower in the radial direction is compared
with the extrapolated track impact point position provided by the telescope. The distribution of the
difference in the Y coordinate, $\Delta Y$, together with the Gaussian fit, are shown in Figure~\ref{fig:position_2}. 
Since the telescope position resolution is much better than that of the LumiCal, the resolution of the
latter is obtained from the standard deviation of the fit of $505 \pm 10~\mu m$.

\subsection{Moli\`ere radius results}
In order to construct the average transverse energy  distribution, the radial energy
distribution for each event had to be shifted to the same particle impact point  to take into account the beam profile.
All events were used to build the energy distribution for each distance, ($E_{m}$), in pads ($m$), from the shower core ($m = 0$).
Using this procedure we can get the average transverse energy  distribution for each configuration.

\begin{figure}[h!]
\begin{center}
\subfloat[\label{fig:MR_1}]{\includegraphics[width=.45\textwidth]{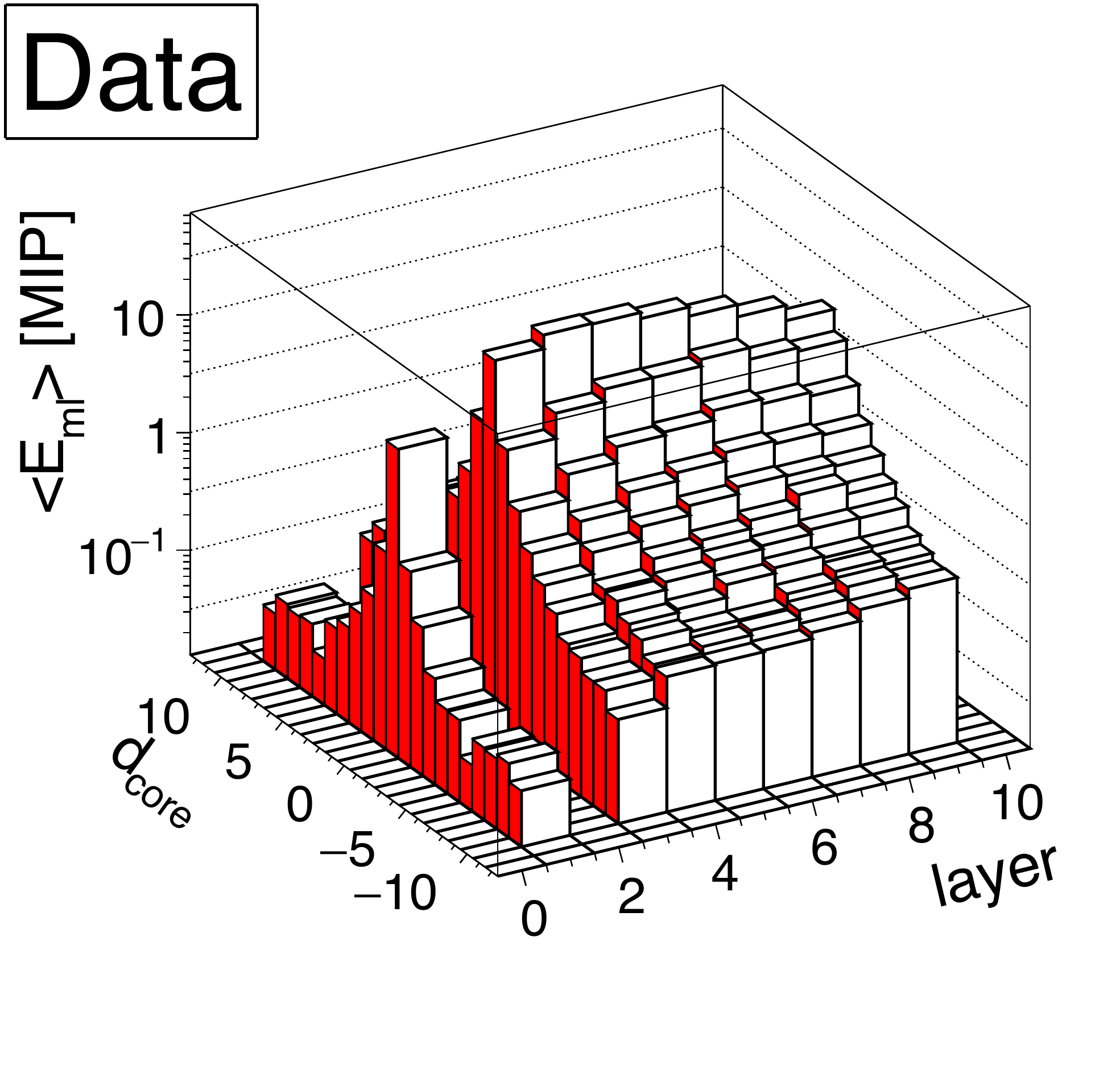}}
\hfill
\subfloat[\label{fig:MR_2}]{\includegraphics[width=.45\textwidth]{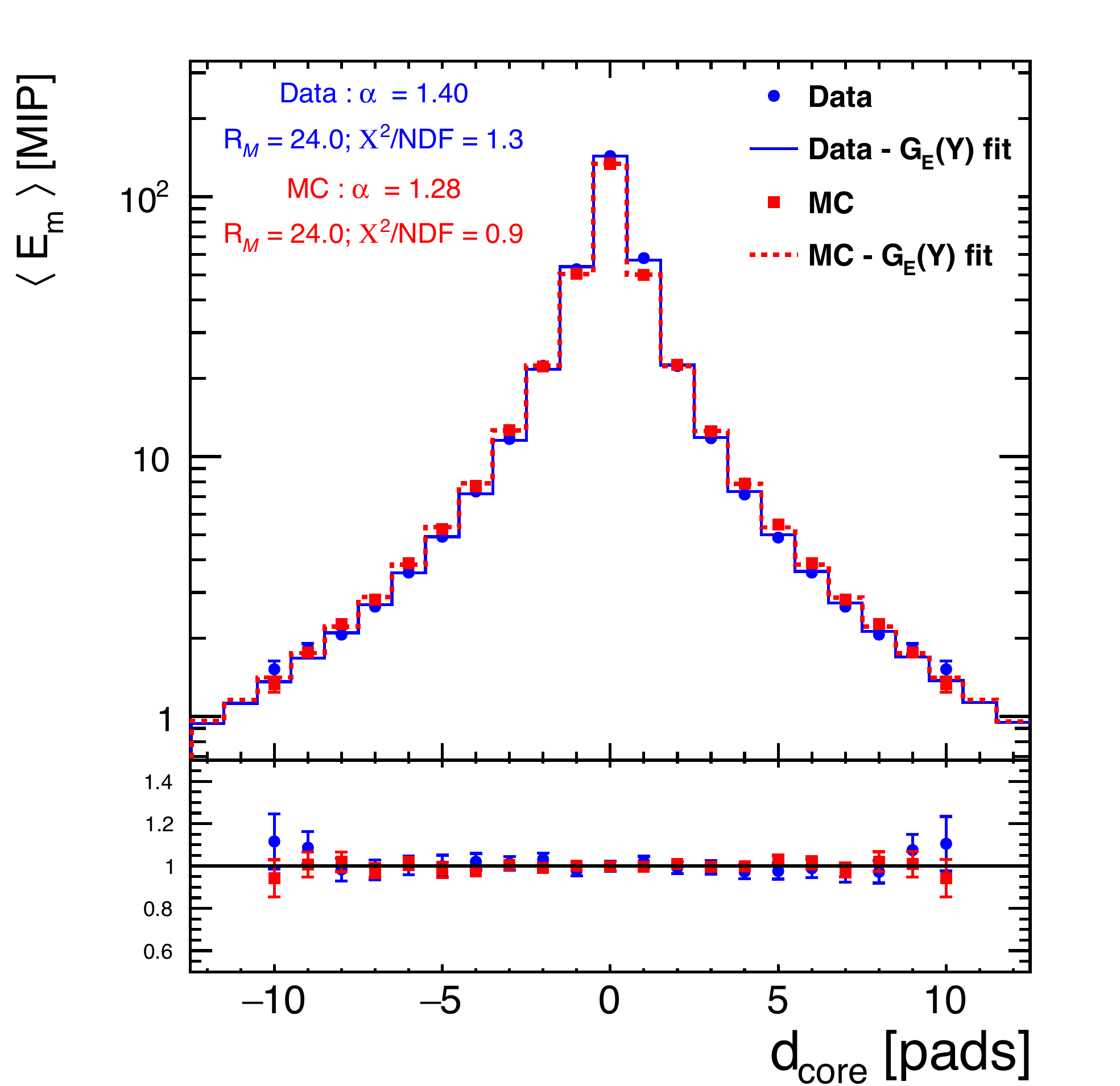}}
\caption{(a) A lego plot of $< E_{ml} >$, as a function of $d_{core}$ in units of pads, for
each layer from the beam test data. (b)  The shower transverse profile $< E_m >$, as a function of $d_{core}$ in units of pads, of
the sum of all three configurations from beam test data and the MC simulation, after symmetry corrections and fit. For the data (blue) and the MC (red).}
\label{fig:MR}
\end{center}
\end{figure}

The three configurations, when properly combined, allow to follow the development of the shower in more detail than each configuration separately and in steps of 1 radiation length.
The average energy deposition in each radial distance from the shower core per layer is denoted by $< E_{ml} >$.
The variable $< E_{ml} >$ as a function of the distance from the shower core, $d_{core}$, is plotted for each layer and presented in
Figure~\ref{fig:MR_1}. 
In order to build the shower transverse profile for all measured layers, the average energy
deposited, $< E_m >$, is constructed as the sum of $< E_{ml} >$ over all layers. 
The shower transverse profile, expressed as the distance from the shower core, $d_{core}$ in units of pads, is presented
in Figure~\ref{fig:MR_2}.

On average, only 10\% of the deposited energy lies outside an infinite long cylinder with a radius of $R_\mathcal{M}$ : 
\small 
\begin{equation} \label{eq:MR_1}
0.9 = \frac {E_{r <  R_\mathcal{M} }}{E_{total}} = \frac{ \int_{0}^{2\pi } r d\varphi \int_{0}^{R_\mathcal{M}}F_{E} ( r )  dr }{ \int_{0}^{2\pi } r d\varphi \int_{0}^{\infty}F_{E} ( r )  dr} . 
\end{equation}
\normalsize
The LumiCal pads are long (strip like) and act like one dimensional integrators making it impossible to
directly access the form of $F_E (r)$. Neglecting the sagitta of the pads, the energy density in the Y direction can be expressed as

\small 
\begin{equation}
G_{E}(y) =\int_{X_{min}}^{X_{max}} F_E(\sqrt{x^2+y^2})dx        .  
\end{equation}
\normalsize
Where the range $(X_{min}, X_{max})$ is defined by the sensor geometry which corresponds to two sectors.
The form of $G_{E}(y)$ is determined by integrating $F_{E}(\sqrt{x^2+y^2} )$ over $x$. 
Thus, by fitting $G_{E}(y)$ to the average energy transverse distribution we gain access to the parameters of $F_{E}(r)$.
The function used for  $F_{E}(r)$ is a Gaussian for the core part and a form inspired by the
Grindhammer-Peters parametrisation~\cite{Grindhammer,Grindhammer1}  to account for the tails,

 \small 
 \begin{equation} \label{eq:MR_FrFinel}
F ( r ) =  (A_C)e^{-(\frac{r}{R_C})^2} + ( A_T )\frac{2r^{\alpha}R_T^2}{ (r^2 + R_T^2 )^2 }       . 
\end{equation}
\normalsize
The results of the numerical fit and calculation give the effective Moli\`ere radius of the configurations used in this test beam to be 24.0 $\pm$ 0.6 (stat.) $\pm$ 1.5 (syst.) mm.
The results of the numerical fit and calculation for the MC simulation give the same outcome of 24.0 $\pm$ 0.6 (stat.).

\section{Summary}

For the first time a multi-plane operation of a prototype of a LumiCal was carried out. 
The development of the electromagnetic shower was investigated and shown to be well described by a GEANT4 Monte Carlo simulation. 
The position resolution for 5 GeV electrons was measured to be $505 \pm 10~\mu m$ and the effective  Moli\`ere radius of the
configurations used was determined to be 24.0 $\pm$ 0.6 (stat.) $\pm$ 1.5 (syst.) mm.
The full description of the test beam and the results can be found at Ref.~\cite{cite:2014}. 

The relatively large value for the Moli\`ere radius obtained in the measurement is due to the large space between the layers creating big air gaps.
To meet the LumiCal design performance, the sensor module thickness was reduced~\cite{Borysov} from 3~mm to less than 1~mm, the size of the gap between absorbers layers.
Carbon fiber supporting structure provided mechanical stability, and thin Kaptons layers for high voltage and fan-out, used to form the new sensor module.
The new module total thickness was below $750~\mu m$, four modules were successfully tested at DESY during 2015, and four additional modules were used during the test in August 2016 including one assembled with TAB bonding technology.
Analysis of 2016 data is ongoing, and preliminary results show that the shower in the transverse plane is significantly narrower. 

\section{Acknowledgments}
This work is partly supported by the Israel Science Foundation (ISF), the Israel German Foundation (GIF), the I-CORE program of
the Israel Planning and Budgeting Committee. This project has received funding from the European
Union's Horizon 2020 Research and Innovation programme under Grant Agreement no. 654168.


\begin{thebibliography}{99}
\bibitem{cite:fcal_jinst}
H. Abramowicz et al.,  [FCAL Collaboration], {\it Forward Instrumentation for ILC Detectors}, {JINST} \textbf{5} (2010) 12002.
\bibitem{cite:TB11} H. Abramowicz et al., [FCAL Collaboration], {\it Performance of fully instrumented detector planes of the forward calorimeter of a Linear Collider detector, } JINST {\bf 10} (2015) P05009.

\bibitem{mimosa26} J. Baudot et al., {\it First test results of MIMOSA-26: A fast CMOS sensor with integrated zero suppression and digitized output, } http://inspirehep.net/record/842163?ln=en.

\bibitem{cite:Oron} O. Rosenblat, {\it Uniformity of detector prototypes for instrumentation in the very forward region of future linear colliders,} MSc Thesis, Tel Aviv University.2016.

\bibitem{nuiry}  F.-X. Nuiry,     {\it Collected documents on the FCAL-AIDA precision mechanical infrastructure and tungsten plates.} https://edms.cern.ch/document/1475879/

\bibitem{fcal_readout_board} Sz. Kulis, A. Matoga, M. Idzik, K. Swientek, T. Fiutowski, D. Przyborowski, 
{\it A general purpose multichannel readout system for radiation detectors} JINST {\bf 7} (2012) T01004.

\bibitem{Jakub} J. Moron, {\it Development of novel low-power, submicron CMOS technology based, readout system for luminosity detector in future linear collider,} Ph. D. dissertation, AGH - UST, Cracow, July 2015.

   \bibitem{Grindhammer}
   G.~Grindhammer et al., in Proceedings of the Workshop on Calorimetry for the Supercollider, Tuscaloosa, AL, March 13-17, 1989, 
   edited by R. Donaldson and M.G.D. Gilchriese (World Scientific, Teaneck, NJ, 1989), p. 151.
   
   \bibitem{Grindhammer1}
   G.~Grindhammer, M.~Rudowicz and S.~Peters, 
  {\it  The Parameterized Simulation of Electromagnetic Showers in Homogeneous and Sampling Calorimeters}, hep-ex/0001020.
   
\bibitem{cite:2014} 
H. Abramowicz et al.,  [FCAL Collaboration], {\it Measurement of shower development and its Molière radius with a four-plane LumiCal test set-up,}, arXiv:1705.03885 [physics.ins-det].

\bibitem{Borysov} O.~Borysov [FCAL Collaboration], {\it Progress Report on an Ultra-compact LumiCal,} arXiv:1703.10496 [physics.ins-det].
\end{thebibliography}
\end{document}